\documentclass[aps,showpacs,twocolumn]{revtex4}
\usepackage{graphics}
\begin{document}
\newcommand{\bstfile}{aps} 
\draft
\title{Monte Carlo Studies of the Fundamental Limits of the Intrinsic Hyperpolarizability}
\author{Mark C. Kuzyk and Mark G. Kuzyk}
\address{Department of Physics and Astronomy, Washington State University, Pullman, Washington
99164-2814}
\date{\today}

\begin{abstract}
The off-resonant hyperpolarizability is calculated using the dipole-free sum-over-stats expression from a randomly chosen set of energies and transition dipole moments that are forced to be consistent with the sum rules.  The process is repeated so that the distribution of hyperpolarizabilities can be determined.  We find this distribution to be a cycloid-like function.  In contrast to variational techniques that when applied to the potential energy function yield an intrinsic hyperpolarizability less than 0.71, our Monte Carlo method yields values that approach unity.  While many transition dipole moments are large when the calculated hyperpolarizability is near the fundamental limit, only two excited states dominate the hyperpolarizability - consistent with the three-level ansatz.
\end{abstract}


\maketitle

\section{Introduction}

Large nonlinear-optical susceptibility materials are central for many optical applications such as telecommunications,\cite{wang04.01} three-dimensional nano-photolithography,\cite{cumps99.01,kawat01.01} and new materials development\cite{karot04.01} for novel cancer therapies.\cite{roy03.01}  In the present work, we focus on the molecular hyperpolarizability, which is the microscopic basis of electro-optic switches and frequency doublers.

The fundamental limits of the hyperpolarizability had been calculated using what appeared to be three-level truncation of the sum rules.\cite{kuzyk00.01,kuzyk03.02} Champagne and Kirtman were critical of the accuracy of truncated sum rules.\cite{Champ05.01} In response, the three-level ansatz was introduced, which states that when a quantum system has a hyperpolarizability near the fundamental limit, the system can be described by a three-level model.\cite{kuzyk05.01}  Using a new dipole-free form of the sum-over-states expression for the hyperpolarizability,\cite{kuzyk05.02} Kuzyk showed rigourously that the three-level ansatz does not demand the sum rules to be truncated.\cite{kuzyk06.03} Furthermore, the three-level ansatz with the new dipole free expression was used to calculate the more general fundamental limits of the dispersion of the real and imaginary parts of the hyperpolarizability.\cite{kuzyk06.03} This result can be used to study any phenomena that is based on a second-order response.  More importantly, since the fundamental limit can be determined at any wavelength, it can be used as an absolute standard for comparing hyperpolarizabilities.  In particular, the ratio of the measured hyperpolarizability to the fundamental limit, called the intrinsic hyperpolarizability, can be used to compare the efficacy of any two molecules, independent of their size.

A factor of thirty gap between the best molecules and the fundamental limit has persisted even after three decades of molecular engineering,\cite{kuzyk03.02} and there is no known explanation for this gap.\cite{Tripa04.01} It is not of a fundamental nature: a simple clipped harmonic potential energy function, which can be solved analytically, falls well into the gap.  As such, a numerical optimization technique, which varies the potential energy function in a way that maximizes the intrinsic hyperpolarizability, was introduced to yield a better understanding of the nature of what makes a large hyperpolarizability.\cite{zhou06.01}  In that work, Zhou and coworkers found that an oscillating wavefunction serves to localize the wavefunctions in a way that forces the system to have only two dominant excited states that interact with the ground state.\cite{zhou06.01}   This result showed how the three-level ansatz is a natural way for optimizing the intrinsic hyperpolarizability.  Based on this observation, Zhou and coworkers proposed that molecules with modulation of conjugation, i.e. with wiggly wavefunctions, may be a new paradigm for making molecules with exceptionally large hyperpolarizabilities.

P\'{e}rez Moreno and coworkers reported on a series of molecules with varying degrees of modulation in the conjugated bridge between the donor and acceptor ends of these quasi-one-dimensional molecules.\cite{perez07.01} Molecules with uneven aromaticity in the bridge, as suggested by Zhou's work, were measured to have larger intrinsic hyperpolarizability than those with a smoother bridge.  The best molecule was reported to have a world's record intrinsic hyperpolarizability - 50\% better than the previous best.

Zhou and coworkers continued to study various ways of optimizing the potential energy function and found that modulated conjugation was but one way of getting near the limit.\cite{zhou07.02}  Depending on the starting potential given to the optimization program, the optimized potential can be wiggly or smooth; the wavefunctions for the optimized potential can be strongly overlapping or well separated; and only a few states or many can contribute.  In all cases, however, the numerical value of the intrinsic hyperpolarizability is just below 0.71 - about 30\% below the fundamental limit.  Thus, there appears to be a large number of distinct potential energy functions that yield the same maximum hyperpolarizability.

These results clearly show that while the three-level ansatz appears to accurately the upper bound of the hyperpolarizability, systems that are not well approximated by three-levels can also have an equally large nonlinear response.  But, it is puzzling that all calculated locally maximal values fall exactly 30\% short of the limit.  In the numerical optimization studies, the potential energy function is varied and the transition moments and energies calculated from the resulting wavefunctions.  Could the sum rules allow for transition moments and energies that do not result from such simple potential energy functions?

To answer these questions, we take the approach of varying the transition moments and energies directly while enforcing compliance with the sum rules, circumventing the need for calculating wavefunctions from a potential.  This process is numerically more efficient and faster than solving the eigenvalue problem for a Hamiltonian, and allows us to investigate a much larger parameter space.  The downside is that we cannot easily associate a potential energy function with those matrix elements and energies that yield the largest hyperpolarizability.

\section{Approach}

We express the sum rules in dimensionless form,
\begin{equation}\label{sumRuleNoD}
\sum_{n=0}^{\infty} \left( e_n - \frac {1} {2} \left(e_m + e_p
\right) \right) \xi_{mn}\xi_{np} = \delta_{mp},
\end{equation}
where $e_n = E_{n0}/E_{10}$ and $\xi{np} = x_{np}/x_{10}^{MAX}$.
$E_n$ is the energy of state $n$ and $E_{nm} \equiv E_n - E_m$. $-e
x_{nm}$ is the electric transition dipole moment between state $n$
and $m$; and $x_{10}^{MAX} \equiv \hbar^2 N_e/2mE_{10}$ is the
fundamental limit of the transition moment between the ground and
first excited state, where $m$ is the mass of the electron, $e$ is
the magnitude of the electron charge and $N_e$ the number of
electrons. Equation \ref{sumRuleNoD} represents an infinite number
of equations, one for each distinct value of $m$ and $p$.  We can
therefore refer to each distinct equation by the pair of indices
$(m,p)$.

In the limit of an $(N+1)$-level model (i.e. only $N$ excited states
contribute to the SOS expression for the hyperpolarizability), we can
treat $(e_1, \ldots, e_N)$ and $\xi_{nm}$ for $n,m \leq N$ as
parameters. Note that we label the ground state with the subscript
zero.  We pick the values of these parameters randomly, but
constrained by the sum rules, as follows.

We start by defining the energy levels.  For an ($N+1$)-level
system, we pick $N$ random numbers in the range $0<r<1$ and order
these in a vector in ascending order so that $r_1<r_2\cdots<r_N$. It
is important that $r_1 \neq 0$ since all of the dimensionless energies
are given by $e_n = r_n/r_1$.  Note that $e_0 = 0$ and $e_1=1$.

Once the energies are determined, the sum rules given by Equation
\ref{sumRuleNoD} are used to fix the transition moments, as follows.
Beginning with the sum rule Equation (0,0), we start by picking a
random number, $r$, in the range $-1<r<+1$.  Since each term in the
sum in Equation (0,0) is non-negative, $\left| \xi_{01} \right|^2 e_1
\leq 1$. But, since $e_1 = 1$, this implies that $\left| \xi_{01}
\right|^2 \leq 1$.  We assume that all of the matrix elements of the
dipole operator are real, so we set $\xi_{01} = \xi_{10} = r$.
Equation \ref{sumRuleNoD} with $N$ excited states and $m=p=0$ can
be written as:
\begin{equation}\label{sumRuleNoD0}
\sum_{n=2}^{N} \left( e_n - e_0 \right) \xi_{m0}^2 = 1 - \left( e_1
- e_0 \right) \xi_{10}^2,
\end{equation}
where the numerical value of $\left( e_1 - e_0 \right)
\xi_{10}^2$ is computed from the already-determined values of $e_1$ and $\xi_{10}$.  Note that $e_0 = 0$, so we drop it.  Since Equation \ref{sumRuleNoD0} implies that
\begin{equation}\label{ineq1}
\xi_{20}^2
\leq \left( 1 - e_1 \xi_{10}^2 \right)/e_2,
\end{equation}
we pick a random number
$r'$ in the range $-1<r'<+1$ and get $\xi_{02}$ using the
expression,
\begin{equation}\label{rule2}
\xi_{02} = \xi_{20} = r'\sqrt{\left( 1 - e_1 \xi_{10}^2
\right)/e_2}.
\end{equation}
To get $\xi_{30}$, we evaluate
\begin{equation}\label{rule3}
\xi_{03} = \xi_{30}
= r'' \sqrt{\left( 1 - e_1 \xi_{10}^2 -  e_2 \xi_{20}^2
\right)/e_3},
\end{equation}
were $r''$ is another random number.  Thus, for the
set of random numbers $r, r', r'', \ldots $, we continue the process
until we have the matrix element $\xi_{10}, \xi_{20},\xi_{30},
\ldots \xi_{N0}$.

Next, we use the sum rule equation (1,1).  Using the same procedure
as above, for a random number $-1<s<+1$,
\begin{equation}\label{rule4}
\xi_{12} = \xi_{21} = s
\sqrt{\left( 1 + e_1 \xi_{10}^2 \right)/(e_2-e_1)}.
\end{equation}
For $-1<s'<+1$,
\begin{equation}\label{rule5}
\xi_{13} = \xi_{31} = s' \sqrt{\left( 1 + e_1 \xi_{10}^2 - (e_2 - e_1) \xi_{21}^2 \right)/(e_3-e_1)},
\end{equation}
and so on, until we
get $\xi_{N1}$.  We continue the process for sum rules $(n,n)$, for
all $n$ where $n<N$.  This procedure then yields all of the
off-diagonal transition moments.  Note that we ignore the sum rule
$(N,N)$ because it gives the nonsensical result that the oscillator strength is negative.

With this protocol, transition moments of the form
$\xi_{nm}$ for $m\neq n$ and the energy levels are determined by the
role of dice; but, these parameters are forced to obey the sum
rules.  We then use these transition moments and energy levels to
calculate the dipole-free intrinsic hyperpolarizability,
$\beta_{int}$, which is given by:\cite{kuzyk05.02}
\begin{equation}\label{dipolefreeBeta}
\beta_{int} = \frac {\beta} {\beta_{MAX}} = \left(
 \frac {3} {4} \right)^{3/4} {\sum_{i \neq j} }' \xi_{0i} \xi_{ij}
 \xi_{j0} \left[\frac {1} {e_i e_j} - \frac {2 e_j - e_i} {e_i^2}
 \right] ,
\end{equation}
where $\beta_{MAX}$ is the fundamental limit of the
hyperpolarizability and $\beta$ is the hyperpolarizability.

Note that because the dipole-free expression was derived from the
sum rules, and explicitly eliminates dipole moments of the form
$\xi_{nn}$, we do not need to determine electric dipole moments of
the system's eigenstates to calculate the intrinsic
hyperpolarizability.

\section{Results and Discussion}

Past studies of optimizing the hyperpolarizabilities used the potential energy function as a starting point.\cite{zhou06.01,zhou07.02}  The wavefunctions for a particular potential energy function are determined from the Schrodinger Equation.  These are then used to determine the dipole matrix and energies of the system, which are in turn the input parameters for calculating the hyperpolarizability using the sum-over-states expression.

The potential energy approach yields several avenues for investigating properties of a system that yield the largest hyperpolarizability.  For example, one can arbitrarily vary the potential energy function to determine directly how the shape of a molecule, characterized by the potential, affects the optical response.  Alternatively, one can define a coulomb potential energy function for various arrangements of nuclei to study structure-property relationships.\cite{kuzyk06.02}  The latter approach, while more realistic because it more closely describes real molecules, is more limited since it is restricted to a superposition of a finite number of point sources.  The former case, on the other hand, might lead to prescriptions for making artificial systems such as multiple quantum well layered structures or nano-engineered structures.

A more basic question pertains to the limitations imposed by restricting the Hamiltonian to be of the form of a kinetic energy term and a potential energy function that depends on position.  It is possible that certain
combinations of energy levels and dipole matrices may be derivable only from a more general Hamiltonian that, for example, contains terms such as $\vec{p} \cdot \vec{f}(r) + \vec{f}(r) \cdot \vec{p}$ (i.e. products of functions of the position, $r$, and the momentum $\vec{p}$) but that may never-the-less still obey the sum rules.  Perhaps these more general Hamiltonians fall into classes that each yield a different fundamental limit, but with $\beta_{int} \leq 1$.  Our Monte Carlo method does not discriminate between these different Hamiltonians as long as the sum rules are obeyed.  Alternatively, matrix elements and energies that are consistent with the sum rules may be inconsistent with some additional basic physical principle.  These are interesting questions that will not be explored here.  Rather, we focus on investigating what matrix elements and energy levels lead to an improved nonlinear optical response; and, how these results compare with those obtained from a simple potential energy function, $V(r)$.

\begin{figure}
\includegraphics{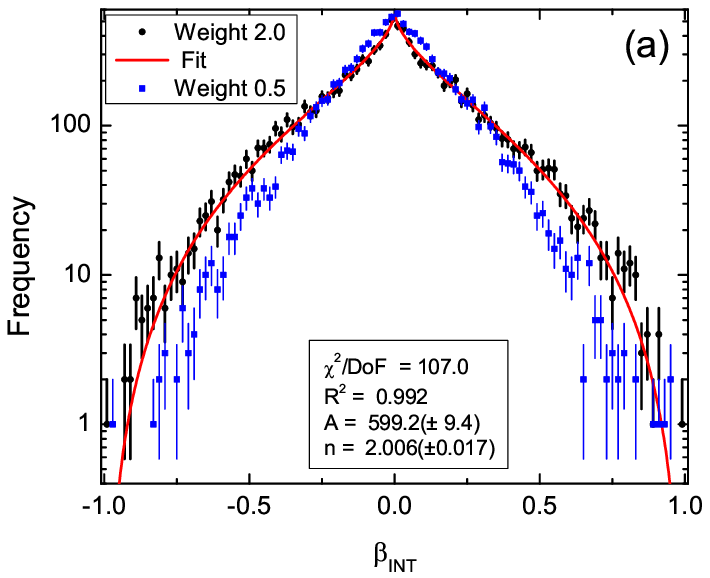}
\includegraphics{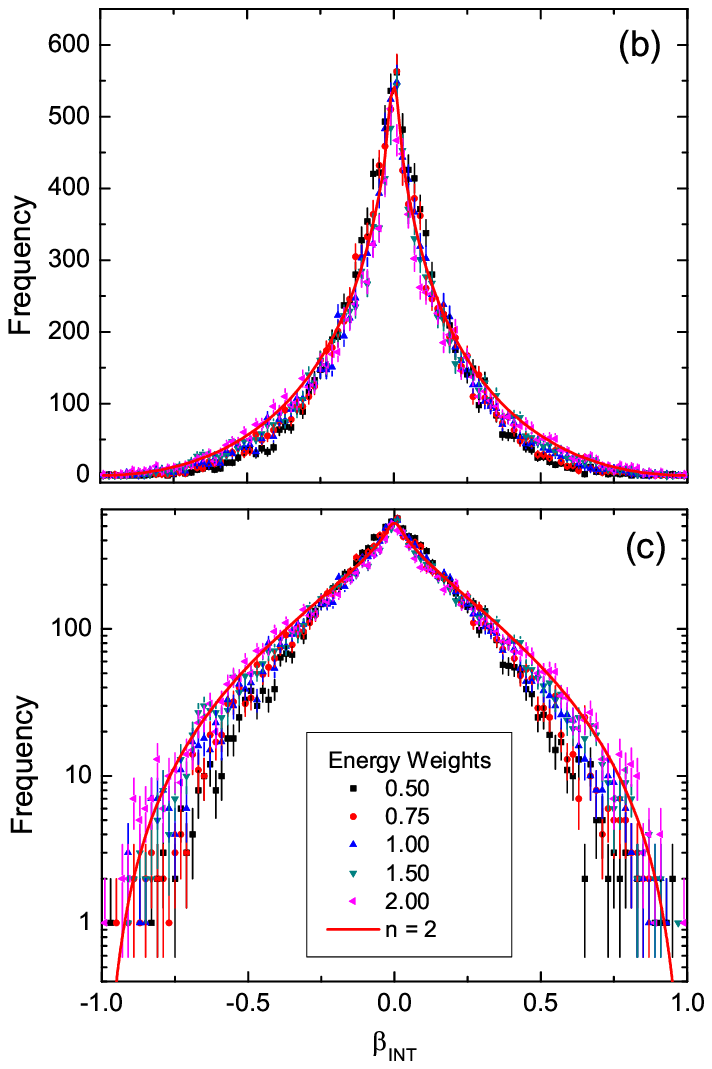}
\caption{Distribution of calculated hyperpolarizabilities for various energy weightings.  Error bars are calculated from the square root of the frequency of observing a particular value of $\beta_{INT}$.}
\label{fig:Distribution}
\end{figure}
Figure \ref{fig:Distribution}a shows the distribution of the calculated intrinsic hyperpolarizabilities for 10,000 runs using a 5-level model.  First we note that all of the calculated values of the intrinsic hyperpolarizability have a magnitude that is less than unity.  This is consistent with the existence of a fundamental limit.  Furthermore, the largest values of $\beta_{INT}$ approach unity.  In contrast, when the potential energy function is varied,\cite{zhou06.01} independent of the starting function, the maximized values converge to about $0.7$.  So, our results suggest that the hyperpolarizabilities attainable from the simple potential energy functions used in past studies may not be as large as is possible based on only the sum rules.  How the sum rules, which are derived from commutation relationships that assume the potential energy function to be dependent on only the positions of the electrons, yield a different upper bound than direct use of the potential energy function is an open question that will be the focus of future studies.  Perhaps such investigations will shed light on why all molecules fall a factor of $10^{3/2}$ below the fundamental limit.\cite{Tripa04.01}

If the energy levels are chosen randomly, they will on average be equally spaced.  To probe a larger parameter space, we define a power law weighting function when randomly choosing the energy level spacing.  Thus, on average, the energy level spacing will be of the form $E^p$.  Figure \ref{fig:Distribution}a shows the result for weighting factors with exponents $p=2$ and $p=0.5$.  The solid curve shows the function
\begin{equation} \label{Eq:distribution}
F = A \left(1 - \beta_{INT}^{1/n} \right)^n ,
\end{equation}
which is fit to the $p=2$ simulation with $n$ and
$A$ as adjustable parameters.  The best fit yields $n=2$.  1.38\% of
the values fall outside the range $\left| \beta_{int} \right|>0.71$.

Figures \ref{fig:Distribution}b and \ref{fig:Distribution}c show the
5-level model results for $p=0.5, 0.75, 1.0, 1.5$, and $2.0$.  The
curve given by Equation \ref{Eq:distribution} with $n=2$ is shown for reference.  Clearly, the
simulations with smaller weighting factors, $p$, yield a narrower
distribution, showing that the number of simulations giving large intrinsic hyperpolarizability
is smaller then the case when $p$ is larger.  This
is consistent with our recent 3-level model analysis that suggests
the second excited state energy should be much higher than the first
in order to optimize the hyperpolarizability,\cite{Tripa04.01,perez07.02} that is, small $E_{10}/E_{20}$ leads to large hyperpolarizability.  Since, on average, $E_{10}/E_{20} = (1/2)^p$ -- the intrinsic hyperpolarizability is larger for larger $p$.

\begin{figure}
\includegraphics{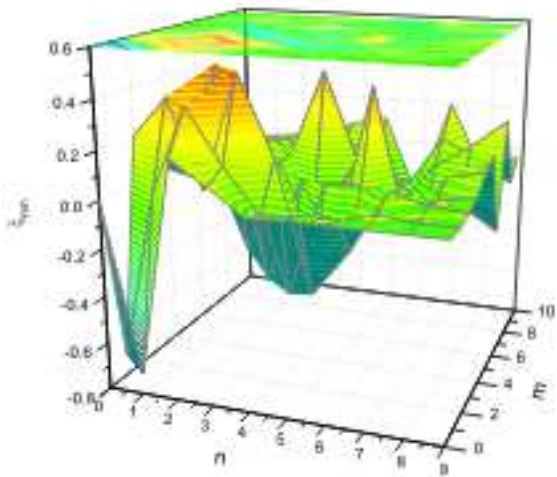}
\caption{The matrix elements that yield the largest value of the
intrinsic hyperpolarizability ($\beta_{int} = 0.9569$) after 100,000
tries for a 10-state model. } \label{fig:MatrixElements10pm}
\end{figure}
To study the character of the states of a system that leads to the
largest nonlinear response, the transition dipole moments and
energies that yield the largest value of the intrinsic
hyperpolarizability from the set of 100,000 random values generated
by our protocol is studied in more detail. Figure
\ref{fig:MatrixElements10pm} shows the matrix elements corresponding
to the largest hyperpolarizability($\beta_{int} = 0.9569$) in the
set for a ten-state model.  For this set, the energies, in
dimensionless units of the form $e_n = (E_n - E_0)/E_1 $ are : 0, 1,
10.08418, 11.16255, 17.29753, 20.40493, 22.37591, 27.98258,
34.93178, and 45.98839.

There are several excited states with substantial transition
moment, though most of the oscillator strength resides in the
lower-lying states. However, in assessing the contribution of each
state to the hyperpolarizability, it is more appropriate to use
Equation \ref{dipolefreeBeta} to define the fractional contribution
from pairs of individual states $n$ and $m$,
\begin{equation}\label{dipolefreeBetaTerms}
\beta_{int}^{n,m} = \left(
 \frac {3} {4} \right)^{3/4} \xi_{0n} \xi_{nm}
 \xi_{m0} \left[\frac {1} {e_n e_m} - \frac {2 e_m - e_n} {e_n^2}
 \right] ,
\end{equation}
\begin{figure}
\includegraphics{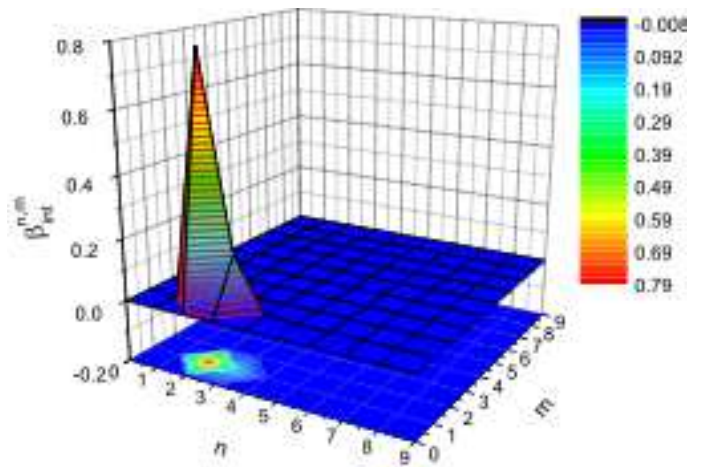}
\caption{The fractional contribution, $\beta_{int}^{n,m}$, for a
10-state model with $\beta_{int} = 0.9569$ (using matrix elements from Figure \ref{fig:MatrixElements10pm}).}
\label{fig:BetaContribute10pm}
\end{figure}
Figure \ref{fig:BetaContribute10pm} shows a plot of the fractional
contribution to the intrinsic hyperpolarizability due to states $n$
and $m$.  Note that $\beta_{int}^{n,m}$, which is calculated from
Equation \ref{dipolefreeBetaTerms}, is not symmetric to interchange
of $n$ and $m$.  However, since $\beta$ is calculated from a sum over
terms of the form $\beta_{int}^{n,m}+\beta_{int}^{m,n}$, the
resulting hyperpolarizability is symmetric upon interchange of the
indices.

Clearly, the two states that contribute most strongly to the
intrinsic hyperpolarizability are states 1 and 2. State 3
contributes, but is small in comparison. All other states are
negligible.  This clearly shows that when the hyperpolarizability is
maximized, the system reduces to a three-state model - consistent with
the three-level ansatz.

It is an interesting exercise to assess this
large-hyperpolarizability ``molecule" using an analysis proposed to
treat complex molecules in terms of parameters determined from
properties of the first two dominant excited
states.\cite{Tripa04.01} In this analysis, the sum rules reduce the
SOS expression in the three-level ansatz to\cite{Tripa04.01}
\begin{equation}\label{f(E)G(X)}
\beta_{INT} = f(E)G(X),
\end{equation}
where $E=E_{10}/E_{20}$, $X=x_{10}/x_{10}^{MAX}$,
\begin{equation}\label{f(E)}
f(E) = \left(1-E \right)^{3/2} \left( E^2 + \frac {3} {2} E + 1
\right)
\end{equation}
and
\begin{equation}\label{G(X)}
G(X) = \sqrt[4]{3} X \sqrt{ \frac {3} {2} \left(1-X^4 \right) } .
\end{equation}
Equation \ref{f(E)G(X)} should be a good approximation to a
molecule whose nonlinear response is close to the limit.

With $E=1/10.08$, we get $f(0.099) = 0.991$.  The transition moment
to the first excited state is given by $X=0.712$, which yields
$G(-0.712)= 0.989$.  Thus, the intrinsic hyperpolarizability $\left|
\beta_{INT} \right|=0.98$, which is in excellent agreement with
$\beta_{INT} = - 0.96$ when all 10 states are used to calculate the
intrinsic hyperpolarizability.  Note that while many transition
moments appear to be substantial (as we can see from Figure
\ref{fig:MatrixElements10pm}), the three-level ansatz yields a good
approximation.

\begin{figure}
\includegraphics{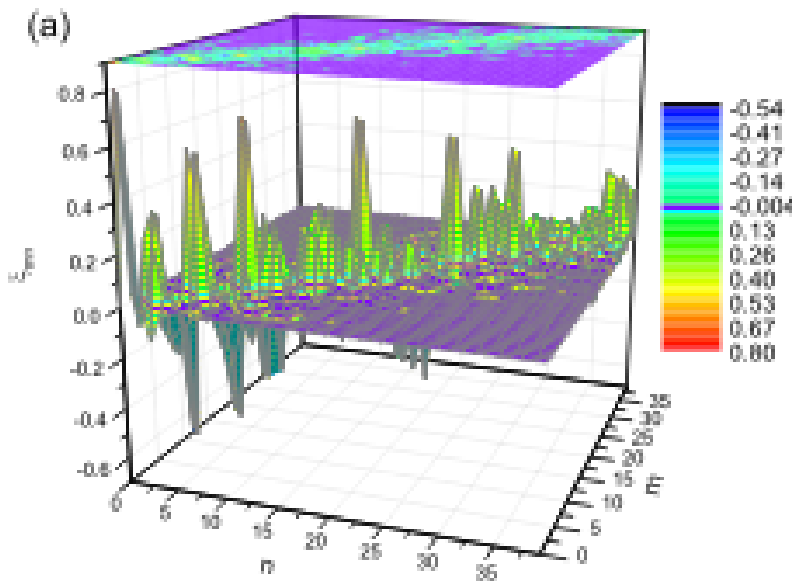}
\includegraphics{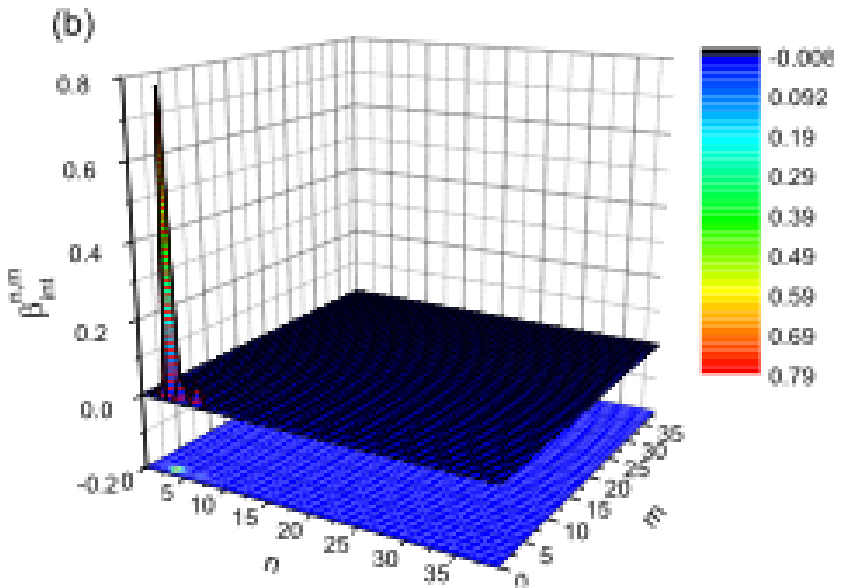}
\caption{(a) The matrix elements that yield the largest value of the
intrinsic hyperpolarizability ($\beta_{int} = 0.9441$) after 100,000
tries for a 40-state model; and (b) the fractional contribution, $\beta_{int}^{n,m}$.} \label{fig:MatrixElements40abs}
\end{figure}
The same pattern holds when more states are included.  Figure \ref{fig:MatrixElements40abs}a shows the matrix elements of a 40-state model that yields $\beta_{int} = 0.9441$.  This is the largest value after 100,000 trials.  As in the 10-level case, we see lots of large matrix elements; but, the fractional contribution to the intrinsic hyperpolarizability, as shown in Figure \ref{fig:MatrixElements40abs}b is dominated by one term corresponding to the first and third excited states.  Again, the three-level ansatz appears to hold.

The relevant state are characterized by $X=0.799$ and $E=0.0433$, which yields $\beta_{INT} = f(0.0433) G(0.799) = 0.998 \times 0.991 = 0.99$.  This is in reasonable agreement with $\beta_{INT}=0.944$, which is obtained when all states are included in the sum.  In contrast to the 10-level model, the 40-level model has $(40/10)^2 = 16$ times more contributions to the hyperpolarizability.  So, when all terms are included there is a larger chance for many smallers terms to add to a significant amount.

In all of our calculations, when the number of states are increased, for example, to 80 states and beyond, the same pattern appears to hold: while many matrix elements are large, only two excited states dominate the intrinsic hyperpolarizability.

\begin{figure}
\includegraphics{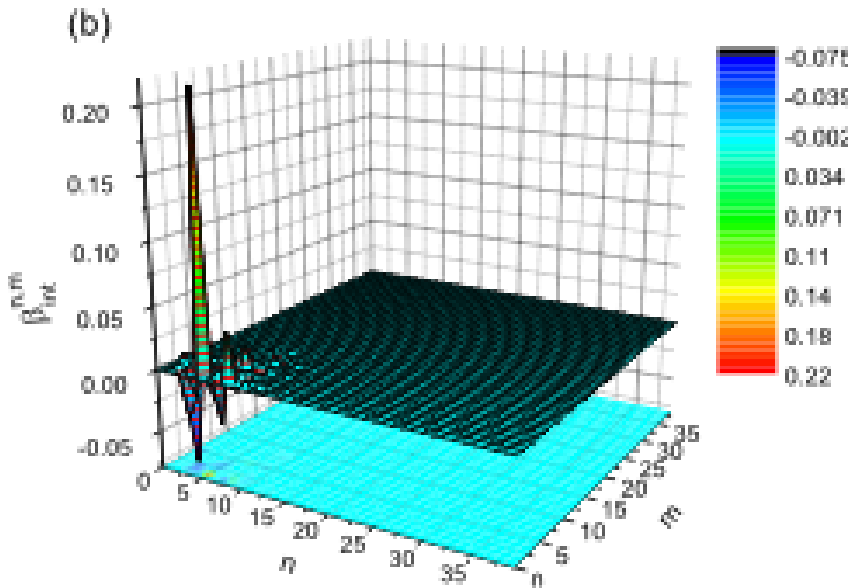}
\includegraphics{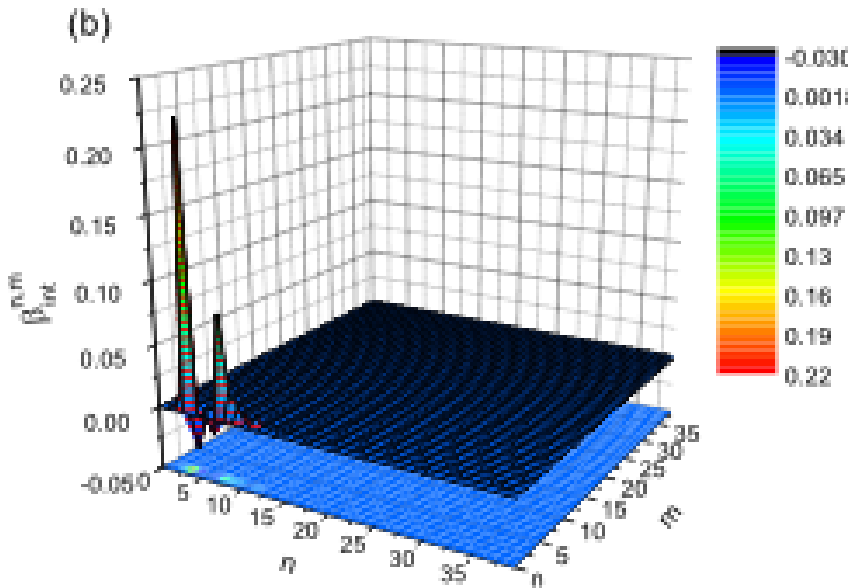}
\caption{The fractional contribution, $\beta_{int}^{n,m}$, for a
40-state model when the intrinsic hyperpolarizability is 0.033 (top) and 0.31 (bottom).}
\label{fig:BetaContribute40pm-point3}
\end{figure}

Figure \ref{fig:BetaContribute40pm-point3} shows the contributions to the hyperpolarizability for two systems, with hyperpolarizabilities of 0.033 and 0.31.  More states contribute as the intrinsic hyperpolarizability gets smaller.  Note that for these two cases, similar to the above results, many transition moments are large, but only a few combinations contribute to $\beta$.

To quantify this observation, we define a term in the SOS expression for $\beta_{INT}^{n,m}$ to contribute substantially if its magnitude is at least 10\% of the term with the largest contribution.  Table \ref{tab:DominantLevels} summarizes the dominant states that correspond to the terms that contribute substantially to the hyperpolarizability. Indeed, the number of states that contribute is typically larger when the intrinsic hyperpolarizability is lower.  We stress that this observation is statistically based.  There are certainly cases where the hyperpolarizability can be large even when many states contribute, as we have found in optimization studies of the potential energy function.\cite{zhou07.02} However, we conclude that when we randomly pick transition moments and energies, the intrinsic hyperpolarizability is {\em usually} small when many states contribute.
\begin{table}
\begin{tabular}{l l l l c}
  \hline
  $\beta_{INT}$ &  & Dominant States &  & Number of Levels \\ \hline
  0.033 &  & 0, 1, 2, 3, 5, 7, and 8 &  & 7 \\
  0.31 &  & 0, 1, 3, 4, and 7 &  & 5 \\
  0.95 &  & 0, 1, and 3 &  & 3 \\
  \hline
\end{tabular}
\caption{States whose contributions are at least 10\% of the largest contribution for the 40-state model shown in Figure \ref{fig:MatrixElements40abs}.}\label{tab:DominantLevels}
\end{table}

It is trivial to show from the sum rules that contributions to
$\beta$ from states of higher energies can dominate over the lower
energy states.  However, since the three-level ansatz implies that
the higher-energy states are not important when the
hyperpolarizability is near the fundamental limit, the higher-lying
states must always cancel each other if any of the terms are
large.

In our protocol, we force the system to be an N-level model.  We can study the effects of the truncation of the sum rules by considering how the hyperpolarizability changes when we add an additional state.  Recall that the diagonal sum rules when truncated to N-levels are of the form,
\begin{equation}\label{sumRuleDiagonal}
\sum_{n=0}^{N-1} e_{np} \xi_{np}\xi_{pn} = 1,
\end{equation}
where we use the shorthand notation $e_{np} = e_n - e_p$.  Equation \ref{sumRuleDiagonal} holds for all $p<N-1$, where we find that $p=N-1$ yields a nonsensical result, so that equation is ignored.

If we add one more state, the diagonal sum rules for $p<N-1$ are of the form,
\begin{equation}\label{sumRuleDiagonal+1}
\sum_{n=0}^{N} e_{np}  \xi_{np}\xi_{pn} = 1.
\end{equation}
With the help of Equation \ref{sumRuleDiagonal}, Equation \ref{sumRuleDiagonal+1} reduces to
\begin{equation}\label{sumRuleDiagonal+1'}
\xi_{N,p}\xi_{p,N} e_{N,p}= 0.
\end{equation}
Assuming that the added state is not degenerate with any of the other $N$ states, this yields
\begin{equation} \label{ZeroXi}
\xi_{N,p} = \xi_{p,N} = 0.
\end{equation}

Next we consider the case with $p=N-1$.  The sum rule,
\begin{equation}\label{sumRuleDiagonal+1"}
\sum_{n=0}^{N} e_{n,N-1} \xi_{N-1,p}\xi_{p,N-1} = 1
\end{equation}
with the fact that $e_{n,N-1} = -e_{N-1,n} < 0$ reduces to,
\begin{equation}\label{sumRuleNewState}
e_{N,N-1} \xi_{N,N-1}\xi_{N-1,N} = 1 + \sum_{n=0}^{N-1} e_{N-1,n}  \xi_{N-1,n}\xi_{n,N-1}
\end{equation}
So, it must be that $\xi_{N,N-1}$ and $e_{N,N-1}$ are non-zero and are related to the values of the transition moments and energies that were determined by the Monte Carlo assignments.

The terms in the dipole-free expression for the hyperpolarizability, given by Equation \ref{dipolefreeBetaTerms}, have numerators of the form $\xi_{0i} \xi_{ij} \xi_{j0}$, where $i \neq j $.  Thus, the only additional terms that contribute to the hyperpolarizability that include the state $N$ are of the form $\xi_{0,N-1} \xi_{N-1,N} \xi_{N,0}$ and $\xi_{0,p} \xi_{p,N} \xi_{N,0}$, where $p<N-1$.  Both of these terms vanish because $\xi_{N0}=0$ according to Equation \ref{ZeroXi}.

If we continue to add more states, we will find that transition moments between these new states and states $N$, $N+1$, etc. will be nonzero; but, the transition moment between the ground state and the newly-added states will vanish because the oscillator strength is depleted by the lowest $N$ states.  The net result is that the additional states do not contribute to the hyperpolarizability.  Since we can continue to add states in this manner, {\em ad infinitum}, to construct a full set of non-truncated sum rules, using truncated sum rules in conjunction with our Monte Carlo method does not appear to lead to any pathologies when calculating $\beta$.

One might ask if it is physical to have a quantum system in which transition moments from the ground state to all states outside of an N-level subspace vanish.  The harmonic oscillator is an example of a system where transitions from the ground state to all excited states beyond the first vanish.  However, it may be that our Monte Carlo simulations may yield transition moments and energies that, while consistent with the sum rules, may not be derivable from a potential energy function.

Champagne and Kirtman had criticized sum-rule truncation, arguing that the resulting limits may be in error.\cite{Champ05.01} The later development of the dipole-free SOS expression for calculating the hypoerpolarizability\cite{kuzyk05.02} lead to a more rigorous argument of why the original calculations, which appeared to truncate the sum rules, did not violate the higher-level sum rules.\cite{kuzyk06.03} A more important concern was the use of the three-level ansatz, which states that the fundamental limits can be calculated by assuming that only two excited states contribute.  Given that the sum rules allow transition moments between excited states to be large, and potentially many states can contribute, one might expect that the calculations of the fundamental limit would underestimate the upper bounds.  This is clearly not the case.  While a system with only two dominant excited states yields the fundamental limit when the ratio of the oscillator strengths to each of the two excited states is chosen appropriately, systems in which many excited states contribute can also reach the fundamental limit.  The implication is that all of the large contributions from transitions between excited states cancel.

\section{Conclusions}

We have shown that when assigning transition moment matrix elements and energies randomly, but under the constraint of the sum rules, values of the intrinsic hyperpolarizability approach unity.  This is in contrast to the intrinsic hyperpolarizability, which when calculated from an optimized potential energy function, falls bellow 0.71.  The distribution of the values of $\beta_{INT}$ is found to be well approximated by a cycloid-like function.  When the energy-level spacing is weighted in a way that forces the lower energy states to have more closely-spaced energies, the distribution falls below the cycloid function, suggesting that more sparse energy-level spacing of the lower levels yields a larger intrinsic hyperpolarizability -- an observation consistent with past analytical studies.

As the intrinsic hyperpolarizability increases, it is found that fewer states contribute even though the transition moments between many states are nonzero and can be large.  Furthermore, when $\beta_{INT}$ is near unity, only three states contribute significantly, and the energy level spacing between the first and second excited states is large compared with the energy difference between the first excited state and the ground state.   This result is consistent with the three-level ansatz.  Furthermore, these matrix elements and energies yield an energy function $f(E)$ and transition moment function $G(X)$ that accurately predicts the hyperpolarizability, which confirms the validity of the type of analysis proposed earlier,\cite{Tripa04.01} and applied to an amylose helix.\cite{perez07.02}

We have also shown that our approach of using a finite number of states, $N$, does not lead to pathologies.  We did so by constructing a state space of dimension $N+1$, whose transition moments and energies satisfy the higher-order sum rules; but, lead to the same hyperpolarizability as for the $N$-dimensional state space.  Since we can construct an infinite-dimensional space by successively adding one state at a time, we conclude that the full set of sum rules are obeyed and the hyperpolarizability remains unchanged from the $N$-dimensional one.

We question whether or not the energy levels and transition moments that are used to calculate the hyperpolarizability when $\beta_{INT} > 0.71$ are derivable from a simple Hamiltonian with a position-dependent potential energy function.  It may very well be the case that, while the sum rules are derived from commutations relationships that assume a potential energy function that depends on position, using only the sum rules as a constraint to determining transition moments and energies may lead to results that belong to a class of more complex Hamiltonians that also satisfy the commutation relationship $[[H,x],x] = \hbar^2/m$.  If this is so, additional constraints may be needed to ensure that such systems are physically reasonable.

{\bf Acknowledgements: } MGK thanks the National Science Foundation
(ECS-0354736) and Wright Paterson Air Force Base for generously
supporting this work.

\clearpage

\end{document}